# IYPt is a highly polar, nonlinear triatomic molecule

Fabio Pichierri

*Department of Applied Chemistry, Graduate School of Engineering, Tohoku University*

*Sendai 980-8579, Japan*

(December 2019)

___________________________________________________________________________

**Abstract.** An explorative quantum chemical study of the triatomic molecule IYPt, its isomers, group 17 congeners, and dimer is carried out. The results indicate that IYPt is a ground-state singlet with a bent geometry and a large electric dipole moment of magnitude 4.8 D. The IPtY isomer also is bent but 33.5 kcal·mol$^{-1}$ higher in energy whereas the isomer with the central iodine atom is not stable a molecule. Furthermore, the calculations indicate that only the dimer of IYPt, made of a (PtY)$_2$ rhombus with the iodine atoms bonded to Y, is characterized by positive vibrational frequencies.

*Keywords*: triatomic molecules, quantum chemistry, DFT, theoretical prediction, Periodic Table

___________________________________________________________________________

1. **Introduction**

In 1869 Dmitry Ivanovich Mendeleev (1834-1907) introduced his original version of the Periodic Table (PT) containing all the 63 elements that were known at the time.[1-4] After 150 years, following the syntheses of transuranic elements (Z ≥ 93) up to oganesson (Z=118), the newest version of the PT represents one of the main pillars of modern science.[5-7] Hence, 2019 was designated by UNESCO as the International Year of the PT (IYPT) of chemical elements and this initiative is supported by IUPAC in partnership with other organizations.[8,9] Nevertheless, researchers have always been inspired by the periodic properties of chemical elements thereby pushing the boundaries of their own disciplines by making insightful theoretical predictions. For instance, in the 1970s elements up to Z=172 were investigated by Fricke et al.[10] while, more recently, Pyykkö[11] suggested possible new molecules for the super heavy elements (SHE) with Z=121-164. The theoretical study[12] of the electronic structure of SHE discovered so far (Z=104-118)[13] is an active field of research that is of paramount importance for both the interpretation and rationalization of nuclear chemistry experiments.

As far as the polyatomic molecules of the lighter chemical elements are concerned, Krüger's original prediction[14] of the heptaatomic LiBeBCNOF molecule named periodane certainly stands out. His study has stimulated further theoretical investigations on periodane,[15] periodane-neon,[16] and the period 3 congener NaMgAlSiPSCl (heavy periodane).[17] Inspired by these studies, we wondered whether the triatomic molecule IYPt could be a stable species on its potential energy surface (PES). Our quest is not too far-fetched given the experimental evidence accumulated so far for the binary combinations of these three chemical elements: (i) the PtY diatomic molecule was identified in the gas-phase[18] while the Pt-Y alloy,[19] following an early suggestion by Nørskov and coworkers[20] based on density functional theory (DFT) calculations, has found useful applications in heterogeneous catalysis;[21] (ii) the structures of both molecular[22] and crystalline[23] YI$_3$ were determined by electron and X-ray diffraction crystallography, respectively, and (iii) the salts of Pt(II) and Pt(IV) iodides also were experimentally characterized.[24]

2. Computational details

All the DFT calculations were performed with the Gaussian 09 software package[25] while pre- and post-processing operations were carried out with the GaussView graphic interface.[26] Geometry optimizations and frequency calculations on IYPt and its isomers employed the B3PW91 hybrid functional[27] in combination with the Def2QZVPP basis set.[28] The choice of the B3PW91 functional was dictated by a recent computational study of several representative inorganic molecules,[29] including transition metals, in which this functional produced the closest structures to those obtained with the *ab initio* coupled-cluster singles and doubles (CCSD) method.[30] Scalar-relativistic effects operative on the core electrons of heavy elements were modelled with the effective core potentials (ECP) associated to the Def2QZVPP basis set while spin-orbit coupling effects were not included. The ECP-28 (Y, I) and ECP-60 (Pt, At) were employed in this study.

3. Results and discussion

Our DFT calculations indicated that any linear arrangement of the three atoms I, Y, and Pt is characterized by two imaginary harmonic vibrational frequencies. On the other hand, the bent geometries of IYPt (130.6°) and IPtY (136.5°) shown in Fig. 1 are characterized by positive vibrational frequencies (vide infra) and therefore both isomers are minima on the PES. In contrast, the bent YIPt isomer with iodine placed between the two transition metals is unstable for it converts to IYPt when the bond angle I-Y-Pt of the initial geometry is smaller than 180°.

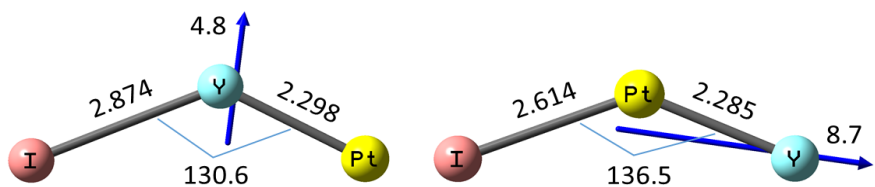

**Fig. 1.** Optimized geometries (B3PW91/Def2QZVPP) of IYPt (left) and its isomer IPtY (right). Bond lengths in angstroms, bond angles in degrees and electric dipole moments in Debye (vectors are scaled x 0.4).

The energy difference ($\Delta E_{iso}$) between the IYPt and IPtY isomers (singlet states) computed at the B3PW91/Def2QZVPP level of theory and inclusive of the corresponding zero-point energy (ZPE) corrections indicates that the former is 33.5 kcal·mol$^{-1}$ lower in energy than the latter. Furthermore, the triplet state of IYPt is 42.9 kcal·mol$^{-1}$ ($\Delta E_{S-T}$) higher in energy than the ground-state singlet. The high-energy isomer IPtY is strongly polar ($\mu$=8.7 D) in comparison to the low-energy one ($\mu$=4.8 D) and their electric dipole moment vectors have different orientations (see Fig. 1), in line with the atomic charges obtained from the natural population analysis.[31] For IYPt the natural charges are −0.56 (I), +1.35 (Y), and −0.79 (Pt) while for IPtY the natural charges are −0.41 (I), −0.57 (Pt), and +0.98 (Y). In both isomers the Y atom is positively charged but its charge increases when it is located at the centre of the molecule. Furthermore, the magnitude of $\mu$ being larger than the accepted threshold value of 2.5 D suggests the formation of dipole-bound anions would be favoured.[32] In this regard, the adiabatic electron affinity (EA) computed for IYPt corresponds to 1.43 eV. For the sake of comparison, the experimental values of EA determined for the homonuclear triatomic molecules $Pt_3$ and $I_3$ are 1.87±0.02 eV and 4.226±0.013 eV, respectively.[33]

In Fig. 2 are displayed selected frontier molecular orbitals of IYPt. The highest-occupied molecular orbital, HOMO, is localized mainly on the Pt-Y bond while the lowest-unoccupied molecular orbital, LUMO, is mainly localized on the central Y atom, in agreement with its positive natural charge (see above). While HOMO-3 has antibonding character, HOMO-7 and HOMO-8 display bonding character for both the I-Y and Y-Pt bonds. Interestingly, HOMO-5 is characterized by a through-space bonding contribution between the iodine and platinum atoms which are separated by the relatively long distance of 4.706 Å. When such distance is reduced by decreasing the bond angle I-Y-Pt from its equilibrium value (130.6°), however, the terminal iodine and platinum atoms do not form a I-Pt covalent bond with each other meaning that the triangular arrangement of these three atoms is not a stable geometry on the PES.

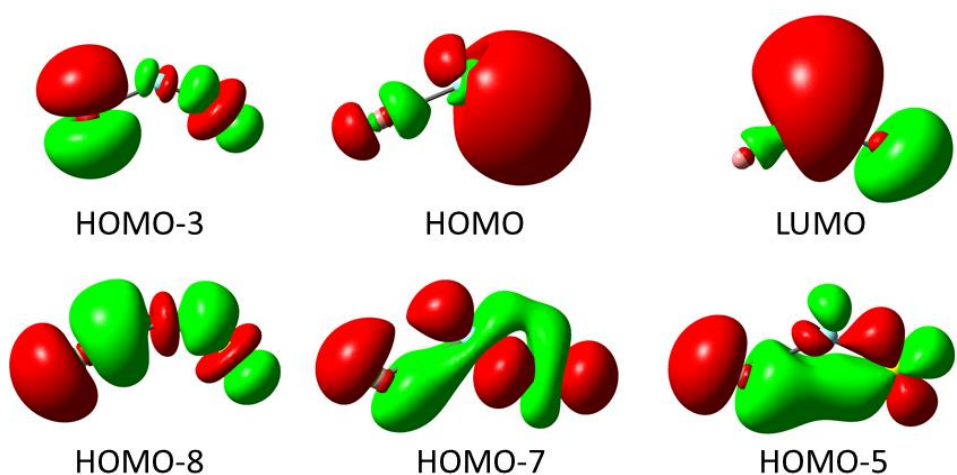

**Fig. 2.** Selected frontier molecular orbitals of IYPt (isovalue=0.02 au).

Next, we investigated the triatomic congeners obtained from the replacement of iodine with the lighter group 17 halogens as well as with the heavier astatine (Z=85), namely XYPt (X=F, Cl, Br, At). The results of our DFT calculations are collected in Table 1 which also includes those for IYPt. As we move down group 17, the Y-Pt distance decreases from 2.328 Å (FYPt) to 2.300 Å (AtYPt) while the bond angle X-Y-Pt increases by about 10°. The molecular polarizability ($\alpha$) increases down group 17 while the magnitude of the electric dipole moment ($\mu$) decreases from 5.4 D (FYPt) to 4.6 D (AtYPt). IYPt has the largest HOMO-LUMO (H-L) gap (4.8 eV) of the family meaning that this molecule is predicted to be the congener with the highest kinetic stability. All the congeners are characterized by adiabatic ionization potentials (IP) above 7.0 eV. The singlet state becomes more stable for the congeners with the heavier halogens, IYPt and AtYPt, while FYPt has the smaller $\Delta E_{S-T}$ value (36.7 kcal·mol$^{-1}$) of the family. On the other hand, the lightest congener FYPt has the largest value of $\Delta E_{iso}$ (67.0 kcal·mol$^{-1}$) which indicates that its isomer FPtY is significantly destabilized.

The frequencies of the harmonic normal modes of vibration are reported in the bottom three rows of Table 1. The bending mode $\nu_1$ is the one with the lowest frequency which decreases three folds on going from FYPt (99 cm$^{-1}$) to AtYPt (32 cm$^{-1}$). The normal modes $\nu_2$ and $\nu_3$ are concerned mainly with the stretching of the X-Y and Y-Pt bonds, respectively. The frequency of these normal modes decreases by less than 50% as the mass of the halogen atom increases along the group.

**Table 1.** Molecular mass and theoretically predicted parameters for the XYPt triatomic molecules (X=F–At).

| Param.[a] | F | Cl | Br | I | At |
|---|---|---|---|---|---|
| Mass (amu) | 302.869 | 318.839 | 362.788 | 410.771 | 494.858 |
| X-Y (Å) | 1.996 | 2.482 | 2.647 | 2.874 | 2.964 |
| Y-Pt (Å) | 2.324 | 2.306 | 2.302 | 2.298 | 2.298 |
| X-Y-Pt (°) | 121.2 | 125.9 | 127.7 | 130.6 | 131.7 |
| α (au) | 100.3 | 110.3 | 117.5 | 133.0 | 142.8 |
| μ (Debye) | 5.5 | 5.2 | 5.1 | 4.8 | 4.6 |
| H-L gap (eV) | 3.19 | 3.33 | 3.37 | 3.40 | 3.38 |
| IP (eV) | 7.40 | 7.55 | 7.56 | 7.26 | 7.13 |
| EA (eV) | 1.22 | 1.35 | 1.39 | 1.43 | 1.44 |
| $\Delta E_{S-T}$ (kcal·mol$^{-1}$) | 36.7 | 39.3 | 39.9 | 42.9 | 40.7 |
| $\Delta E_{iso}$ (kcal·mol$^{-1}$) | 67.0 | 45.3 | 40.3 | 33.5 | 31.6 |
| $\nu_1$ (cm$^{-1}$) | 99 | 61 | 45 | 36 | 32 |
| $\nu_2$ (cm$^{-1}$) | 282 | 277 | 222 | 176 | 152 |
| $\nu_3$ (cm$^{-1}$) | 564 | 343 | 302 | 300 | 299 |

[a] The molecular masses utilized in the calculations are reported here to the third decimal place. The adiabatic ionization potential (IP), adiabatic electron affinity (EA), singlet-triplet energy difference ($\Delta E_{S-T}$), and the energy difference between the XYPt and XPtY (X=F-At) isomers ($\Delta E_{iso}$) are ZPE-corrected values.

The final issue that we wish to investigate here is whether the dimers of IYPt and IPtY are stable species. Our DFT calculations indicated that the D$_{2h}$-symmetric dimer (IYPt)$_2$ shown at the top of Fig. 3, with the iodine atoms bonded to yttrium, is characterized by positive harmonic vibrational frequencies whereas the one at the bottom, with the iodine atoms bonded to platinum, has one imaginary frequency (−29 cm$^{-1}$). An analysis of the simulated vibrational spectrum of (IYPt)$_2$ indicates that the 10-th normal mode (222 cm$^{-1}$) contributes a very intense IR band (192.5 km·mol$^{-1}$) which should be experimentally identifiable. Such vibrational mode corresponds to the oscillatory displacement of the Y atoms along the direction of the I–I internuclear axis. We notice that the Y-Pt bond distance in the dimer is significantly elongated (2.599 Å) with respect to the monomer (2.298 Å) whereas the I-Y distance is not affected by the dimerization. Furthermore, the ZPE-corrected binding energy computed for the dimer is

−88.0 kcal·mol$^{-1}$ while its Gibbs free energy of formation at 298.15 K and 1.0 atm corresponds to −76.5 kcal·mol$^{-1}$ which indicates that the dimerization reaction is exergonic.

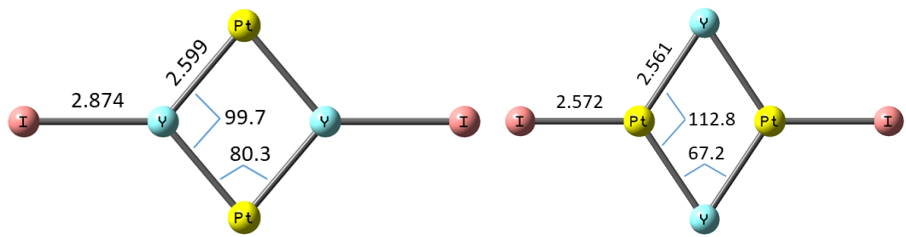

**Fig. 3.** Optimized geometries of the IYPt (left) and IPtY (right) dimers. Bond lengths in angstroms and bond angles in degrees.

As remarked by Krüger in his original paper on periodane,[14] the theoretical predictions of novel inorganic molecules[34] are less frequent than those of organic molecules. This is likely due to the large number of bonding situations encountered in inorganic chemistry which result from the different oxidation states of metals as well as the increasing importance of relativistic effects[35-37] for the heavy elements. We therefore suggest the search for novel inorganic species should start from triatomic molecules which represent the building blocks for the chemical syntheses of complex materials.[38] For instance, heavy triatomics such as HgI$_2$ may find potential applications in the development of nonlinear optical materials[39] while the laser-coolable RaOH molecule is currently employed in the study of molecular fountains for the accurate determination of physical constants.[40] Finally, some time ago, using Hamiltonians made of 2-4 coupled Morse oscillators, we have shown that the increase in molecular complexity on going from tri- to pentaatomic molecules affects the statistical properties of vibrational energy levels.[41] This corresponds to the transition of the nearest-neighbor level spacing distribution (NNLSD) from a Poisson-like to a Wigner-like distribution. Hence, triatomic molecules are ideally suited for testing the modern theories of chemical bonding[42-45] as well as the possible extension of Mendeleev's periodic law from the atomic to the molecular domain.[46]

## 4. Conclusions

Taking inspiration from early theoretical studies of periodane and related compounds,[14-17] we investigated the molecular and electronic structure of the inorganic, triatomic molecule IYPt. Our calculations indicate IYPt in its singlet ground-state is bent ($C_s$-symmetry) while the triplet state is considerably higher in energy (42.9 kcal·mol$^{-1}$). The isomer IPtY also is bent and 33.5 kcal·mol$^{-1}$ higher in energy whereas the isomer with iodine placed in the middle (YIPt) is not a stable species on the PES. The study was extended to the XYPt congeners containing both lighter (X=F, Cl, Br) and heavier (X=At) group 17 elements all of which are nonlinear as well. These neutral, triatomic molecules are significantly polar with dipole moments whose magnitude decreases along the elements of group 17. The dimerization of IYPt yields an inorganic molecule made of a $(PtY)_2$ rhombus with the iodine atoms bonded to the Y atoms located on the opposite corners. When the iodine atoms are bonded to the platinum atoms, however, the resulting molecule which arises from the dimerization of IPtY is not a minimum on the PES. Besides an explicit treatment of relativistic effects, possible extensions of the present work could be devoted to study the replacement of the terminal Pt atom in IYPt with either Pd or Ni atoms as well as investigating the compounds derived from the attachment to IYPt of other elements of the PT. Experimental work is eagerly awaited to validate the present theoretical predictions.

**Declaration of Competing Interest**

The author declares that there is no competing interest.

**Acknowledgments**

The author thanks the Department of Applied Chemistry of the Graduate School of Engineering, Tohoku University, for support.